\documentclass[a4paper,11pt]{article}
\pdfoutput=1
\usepackage{jheppub}
\usepackage{amsmath}
\usepackage{hyperref}
\allowdisplaybreaks
\def\bea#1\eea{\begin{align}#1\end{align}}
\newcommand{\nnu}{\nonumber\\}
\newcommand{\kt}{anti-k$_{\rm T}$}
\newcommand{\bef}{\begin{figure}[hbt]\centering}
\newcommand{\eef}{\end{figure}}

\title{Jet fragmentation functions in proton-proton collisions using soft-collinear effective theory}

\author{Yang-Ting Chien,}
\author{Zhong-Bo Kang,}
\author{Felix Ringer,}
\author{Ivan Vitev}
\author{and Hongxi Xing}

\affiliation{Theoretical Division, 
                   Los Alamos National Laboratory,
                   Los Alamos, NM 87545, USA}

\emailAdd{ytchien@lanl.gov}
\emailAdd{zkang@lanl.gov}
\emailAdd{f.ringer@lanl.gov}
\emailAdd{ivitev@lanl.gov}
\emailAdd{hxing@lanl.gov}

\abstract{The jet fragmentation function describes the longitudinal momentum distribution of hadrons inside a reconstructed jet. We study the jet fragmentation function in proton-proton collisions in the framework of soft-collinear effective theory (SCET). We find that, up to power corrections, the jet fragmentation function can be expressed as the ratio of the fragmenting jet function and the unmeasured jet function. Using renormalization group techniques, we are able to resum large logarithms of jet radii $R$ in the perturbative expansion of the cross section. We use our theoretical formalism to describe the jet fragmentation functions for {\it light} hadron and {\it heavy} meson production measured at the Large Hadron Collider (LHC). Our calculations agree very well with the experimental data for the light hadron production. On the other hand, although our calculations for the heavy meson production inside jets are consistent with the {\sc PYTHIA} simulation, they fail to describe the LHC data. We find that the jet fragmentation function for heavy meson production is very sensitive to the gluon-to-heavy-meson fragmentation function.}

\begin{document}
\maketitle

\section{Introduction}
\label{sec:intro}
Collimated jets of hadrons are a dominant feature of high energy particle interactions, especially at the current highest energy hadron collider, the Large Hadron Collider (LHC), where jets are abundantly produced. The internal structure of these jets has become an important tool to test the fundamental properties of Quantum Chromodynamics (QCD), and to search for new physics beyond the Standard Model~\cite{Altheimer:2013yza,Adams:2015hiv}. Needless to say,  a good understanding of jet substructure allows deeper insights into QCD dynamics and serves as a prerequisite for further progress.

One of the jet substructure observables proposed and explored in more detail recently is the jet fragmentation function, which describes the longitudinal momentum distribution of hadrons inside a reconstructed jet~\cite{Procura:2009vm,Liu:2010ng,Jain:2011xz,Jain:2011iu,Procura:2011aq,Bauer:2013bza,Cacciari:2012mu,Arleo:2013tya,Ritzmann:2014mka,Baumgart:2014upa,Kaufmann:2015hma}. Experimental studies on hadron distribution inside jets have been pioneered at the Tevatron~\cite{Abe:1990ii} in the 1990s. More recently, both the ATLAS and the CMS collaborations have measured the distributions of  {\it light} hadron~\cite{Aad:2011sc,Aad:2014wha,Chatrchyan:2012gw,Chatrchyan:2014ava}  and {\it heavy} meson~\cite{Aad:2011td} production inside jets at the LHC. The jet fragmentation function is an interesting and important observable:  since it probes the hadron fragmentation at a more differential level, it can reveal detailed information about the jet dynamics involved in producing the identified hadron. At the same time, it can provide further information about the non-perturbative hadronization encoded in the standard fragmentation functions. One might even gain insight into the nontrivial spin correlation through the study of azimuthal distribution of the hadron inside jets~\cite{Yuan:2007nd,D'Alesio:2010am,D'Alesio:2011mc,Aschenauer:2013woa,Aschenauer:2015eha}.

Since gluon jets are much more abundant in proton-proton collisions at high energy hadron colliders, jet fragmentation functions should be more sensitive to gluon fragmentation. We will show that this is  the case especially for heavy meson production inside jets. This situation is very different from the $e^+e^-\to h\, X$ and $e\,p\to e\,h\,X$ processes, where the gluon fragmentation function does not enter at leading-order in the perturbative calculation and, thus, can only be probed through QCD evolution or higher-order radiative corrections.

There is also strong motivation to study the jet fragmentation function in heavy ion collisions at high energies, where hot and dense QCD medium -- the quark-gluon plasma -- is  produced. By comparing the jet fragmentation function measured in ultra-relativistic nucleus collisions and the one in proton-proton collisions, one can understand how the presence of the strongly interacting medium produced in heavy ion collisions modifies the hadron distributions inside jets. Understanding the light and heavy flavor dynamics in the medium will help further determine the precise properties of the QGP. For recent experimental measurements of the jet fragmentation function in heavy ion collisions at the LHC, see~\cite{Chatrchyan:2012gw,Chatrchyan:2014ava,Aad:2014wha}. For some theoretical work along this direction, see \cite{Perez-Ramos:2014mna,Mehtar-Tani:2014yea,Casalderrey-Solana:2014bpa}.

In this paper, we study the jet fragmentation function in proton-proton collisions using soft-collinear effective theory (SCET)~\cite{Bauer:2000ew,Bauer:2000yr,Bauer:2001ct,Bauer:2001yt,Bauer:2002nz}. 
Previously, in~\cite{Arleo:2013tya,Kaufmann:2015hma} a full next-to-leading order (NLO) calculation was performed. Closely related work with emphasis on heavy flavor was also recently presented in \cite{jet_frag_angularity,Mehen:2016xly}. As we will show below, within SCET the hadron distribution inside jets is governed by the ratio of two quantities: the fragmenting jet function (FJF) ${\mathcal G}_{i}^{h}(\omega, R, z, \mu)$ introduced and studied in~\cite{Procura:2009vm,Liu:2010ng,Jain:2011xz,Jain:2011iu,Procura:2011aq,Bauer:2013bza}, and the unmeasured jet function $J^i(\omega, R, \mu)$ introduced in~\cite{Ellis:2010rwa}. Here, $i$ is the parton that initiates the jet with energy $\omega$ and  radius $R$, while $z$ is the fraction of the jet momentum carried by the identified hadron $h$. The FJF ${\mathcal G}_{i}^{h}(\omega, R, z, \mu)$ can be further written as a convolution of perturbatively calculable Wilson coefficients ${\mathcal J}_{ij}$ and the fragmentation functions $D^h_j(z, \mu)$. Using the renormalization group techniques, we are able to simultaneously resum logarithms of the form $\ln R$ and $\ln(1-z)$, which have a significant numerical impact. Such resummations were not addressed previously in the fixed NLO calculation of~\cite{Kaufmann:2015hma}. We use the formalism to describe the experimental data at the LHC for the distribution of {\it light} hadron and {\it heavy} meson production inside jets. The study of the jet fragmentation function in heavy ion collisions using SCET will be performed in a forthcoming paper~\cite{jet_frag}. Some of the input for this calculation, such as the final-state in-medium splitting functions~\cite{Ovanesyan:2011kn} and medium-modified fragmentation functions applied to leading hadron production~\cite{Kang:2014xsa,Chien:2015vja}, are already available.  

Here, we would like to remind the readers that, although the jet fragmentation function and the fragmenting jet function look very similar, they have different meanings. It is important to understand their differences and relations since they appear throughout the entire paper. The jet fragmentation function is an experimental observable describing the distribution of hadrons inside jets. On the other hand, the fragmenting jet function is a theoretical quantity which enters the factorized expression in the calculation of the jet fragmentation function. See Sec.~\ref{sec:theory} for more details.

The rest of the paper is organized as follows. In Sec.~\ref{sec:theory}, we first provide the definition of the jet fragmentation function. We then derive a factorized expression for the jet fragmentation function, which involves the FJF and the unmeasured jet function. We give the matching coefficients for the FJF to be convolved with the standard fragmentation functions, and in particular for jets reconstructed using the anti-k$_{\rm T}$ jet algorithm, which is used in almost all  jet reconstruction at the LHC. We collect the detailed derivations of the matching coefficients in the Appendix Sec.~\ref{sec:appendix}. In Sec.~\ref{sec:phenomenology}, we present the numerical results of our calculations for {\it light} hadron and {\it heavy} meson production inside jets and compare with the experimental data at the LHC. We also explore the theoretical uncertainty,  the sensitivity of the observable to the jet algorithm (either cone or \kt),  and  the radius dependence. We summarize our paper in Sec.~\ref{sec:summary}.

\section{Jet fragmentation function}
\label{sec:theory}

In this section we give the definition of the jet fragmentation function and calculate it using the factorized expression in SCET. The evaluation involves the fragmenting jet function ${\mathcal G}_{i}^h(\omega, z,R,\mu)$, and we provide the Wilson coefficients ${\mathcal J}_{ij}$ to be convolved with the fragmentation function $D_{j}^{h}(z, \mu)$. We give the results for jets reconstructed using cone and \kt~algorithms, as ${\mathcal J}_{ij}$ depends on the jet algorithm. The results for cone jets are available in \cite{Procura:2011aq}, while those for \kt~jets were first written down in the appendix of \cite{Waalewijn:2012sv}. We provide the detailed derivations of ${\mathcal J}_{ij}$ for \kt~ jets in the appendix, and the results are consistent with \cite{Waalewijn:2012sv}. 

\subsection{Observable and factorized expression}
\label{factorization}
The jet fragmentation function $F(z, p_T)$ describes the longitudinal momentum distribution of hadrons inside a reconstructed jet. We will compare our calculations with the jet fragmentation functions measured in proton-proton collisions, $p+p\to ({\rm jet~with~}\, h)+X$. Here, $F(z, p_T)$ is defined as follows,
\bea
F(z, p_T) = \frac{d\sigma^h}{dy dp_T dz}\Big/\frac{d\sigma}{dy dp_T},
\eea
where $d\sigma^h/dy dp_T dz$ and $d\sigma/dy dp_T$ are the differential cross sections of jets with and without the reconstruction of the hadron $h$ in the jet. Here, $y$ and $p_T$ are the jet rapidity and transverse momentum. $z$ is the fraction of the jet transverse momentum carried by the hadron, $z\equiv p_T^h / p_T$, with $p_T^h$ the transverse momentum of the hadron. Jets are reconstructed using either the cone or the \kt~ algorithm with the jet radius $R$, and the $R$-dependence is suppressed in the expression for $F(z, p_T)$. As we will see, jet fragmentation functions will be different for jets reconstructed using different jet algorithms.

\bef
\includegraphics[width=2.2in]{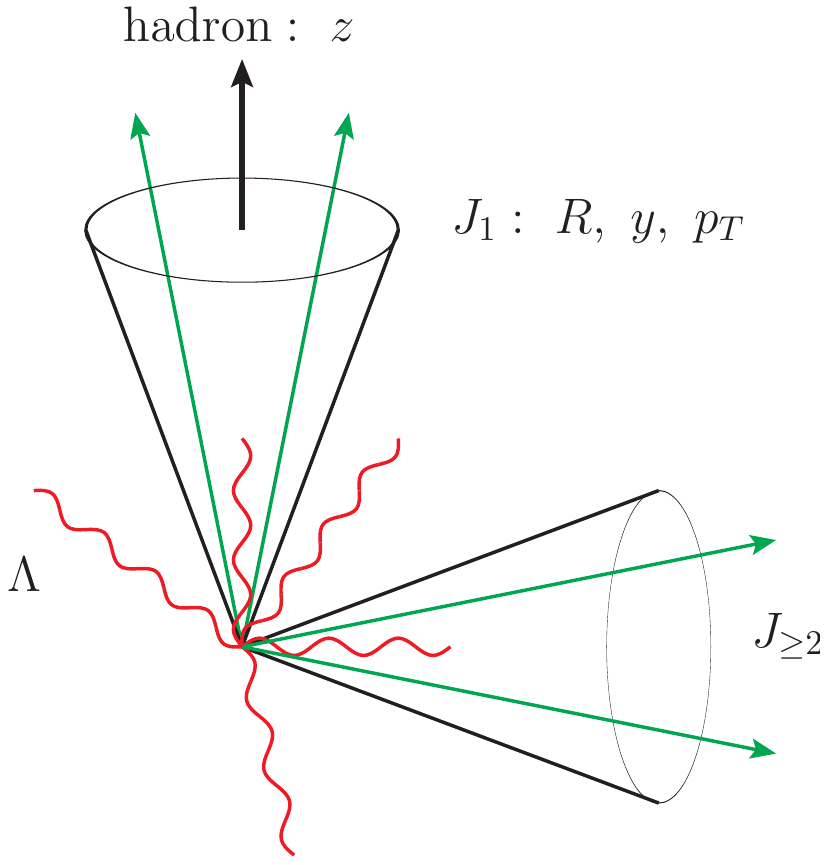}
\caption{Illustration of the $N$-jet production in $e^+e^-$ collisions, where a hadron is measured in the jet labeled by $J_1$ with rapidity $y$ and transverse momentum $p_T$. $z$ is the fraction of the jet momentum carried by the hadron. Jets are reconstructed using a jet algorithm with radius $R$. We impose an energy cutoff $\Lambda$ outside the jets to ensure the $N$-jet configuration. $\Lambda$ is a low energy scale constraining the soft radiation (red lines). The green lines represent the collinear splittings.}
\label{fig:n-jet}
\eef

Because the contribution from the soft radiation to the longitudinal momentum is power suppressed \cite{Chien:2014nsa}, it suffices to illustrate the SCET factorized expression for the jet fragmentation function in $e^+e^-$ collisions (Fig.~\ref{fig:n-jet}). Following \cite{Procura:2009vm,Procura:2011aq,Ellis:2010rwa,Chien:2014nsa,Chien:2015hda,Baumgart:2014upa}, the differential cross section for $N$-jet production with the jet $p_{T_i}$ and $y_i$, the hadron $h$ inside one jet (labeled by 1), and the energy cutoff $\Lambda$ outside all the jets can be written as follows,
\bea
\frac{d\sigma^h}{dy_i dp_{T_i} dz} = H(y_i, p_{T_i}, \mu) \, {\mathcal G}_{\omega_1}^h(z, \mu)
J_{\omega_2}(\mu)\cdots J_{\omega_N}(\mu) S_{n_1 n_2 \cdots n_N} (\Lambda, \mu) + {\mathcal O}\left(\frac{\Lambda}{Q}\right) + {\mathcal O}(R),
\label{eq:factorization}
\eea
where $H(y_i, p_{T_i}, \mu)$ is the hard function describing the short-distance production of the $N$ jets with rapidities $y_i$ and momenta $p_{T_i}$. 
$S_{n_1 n_2 \cdots n_N} (\Lambda, \mu)$ is the soft function with $N$ soft Wilson lines along the jet directions. The energy cutoff $\Lambda$ outside the jets is imposed to ensure the $N$-jet configuration. The hadron $h$ measured inside jet 1 is described by the FJF ${\mathcal G}_{\omega_1}^h(z, \mu)$, with the jet radius $R$ suppressed. $J_{\omega_{i}}(\mu)$ (for $i=2,\cdots, N$) are the unmeasured jet functions introduced in \cite{Ellis:2010rwa} \footnote{The full logarithmic structure of the unmeasured jet function at ${\cal O}(\alpha_s^2)$ is derived in \cite{chien:2015cka}.}, with $\omega_i$ representing the large light-cone component of the jet momentum and $\omega_i = 2p_{T_i}$ in the frame where the jet is 
in the transverse direction. The factorized expression is valid for collimated jets up to power corrections of the type $\Lambda/Q$ or $R$. 

On the other hand, the differential cross section for $N$-jet production is given by
\bea
\frac{d\sigma}{dy_{i} dp_{T_i}} = H(y_i, p_{T_i},\mu) J_{\omega_1}(\mu)\cdots J_{\omega_N}(\mu)  S_{n_1 n_2 \cdots n_N} (\Lambda, \mu) + {\mathcal O}\left(\frac{\Lambda}{Q}\right) + {\mathcal O}(R),
\label{eq:jet}
\eea
with the same hard function $H(y_i, p_{T_i},\mu)$, soft function $S_{n_1 n_2 \cdots n_N} (\Lambda, \mu)$, and unmeasured jet functions $J_{\omega_i}(\mu)$ with $i=2,\cdots, N$. The only difference is that ${\mathcal G}_{\omega_1}^h(z, \mu)$ in Eq.~\eqref{eq:factorization} is replaced by the unmeasured jet function $J_{\omega_1}(\mu)$ in Eq.~\eqref{eq:jet} since we do not measure the hadron. The distribution of the hadron $h$ inside jet 1 then becomes,
\bea
F_{\omega_1}(z, p_{T_i}) = \frac{d\sigma^h}{dy_i dp_{T_i} dz}\Big/\frac{d\sigma}{dy_{i} dp_{T_i}} = \frac{{\mathcal G}_{\omega_1}^h(z, \mu)}{J_{\omega_1}(\mu)}.
\label{eq:ffs}
\eea
All the hard, soft and unmeasured jet functions (except for jet 1) cancel in the ratio. Taking the average over the jet production cross section, with proper phase space (PS) cuts on both jet rapidity $y$ and transverse momentum $p_T$, e.g. the rapidity interval and the width of the $p_T$ bin, the jet fragmentation function $F(z, p_T)$ becomes
\bea
F(z, p_T) = \frac{1}{\sigma_{\rm total}} \sum_{i=q, g} \int_{\rm PS} dy\, dp_{T'} \frac{d\sigma^i}{dy\,dp_{T'}} \frac{{\mathcal G}_i^h(\omega, R, z, \mu)}{J^i(\omega, R, \mu)},
\label{eq:master}
\eea
where $d\sigma^i/dy\,dp_{T'}$ is the cross section to produce the jet initiated by parton $i$, and we have written out explicitly the arguments for both the FJF ${\mathcal G}_i^h(\omega, R, z, \mu)$ and the unmeasured jet function $J^i(\omega, R, \mu)$.

In the next subsection we will provide explicit expressions for the fragmenting jet function ${\mathcal G}_i^h(\omega, R, z, \mu)$ and the unmeasured jet function $J^i(\omega, R, \mu)$. Here it is instructive to point out that ${\mathcal G}_i^h(\omega, R, z, \mu)$ and $J^i(\omega, R, \mu)$ have the same renormalization group (RG) evolution \cite{Jain:2011xz,Procura:2011aq,Ellis:2010rwa} and the ratio ${\mathcal G}_i^h(\omega, R, z, \mu)/J^i(\omega, R, \mu)$ is renormalization group invariant, with possibly different characteristic scales for ${\mathcal G}_i^h$ and $J^i$.

\subsection{Unmeasured jet function}
For convenience, we provide all the relevant results for the unmeasured jet function $J^i(\omega, R, \mu)$. At ${\cal O}(\alpha_s)$ \cite{Ellis:2010rwa},
\bea
J^q(\omega, R, \mu) &= 1+ \frac{\alpha_s}{\pi} C_F \left[L^2 - \frac{3}{2}L + d_J^{q, \rm alg} \right],
\label{eq:Jq}
\\
J^g(\omega, R, \mu) &= 1+ \frac{\alpha_s}{\pi} C_A \left[L^2 - \frac{\beta_0}{2C_A} L + d_J^{g,\rm alg}\right],
\label{eq:Jg}
\eea
where
\bea
L = \ln\frac{\omega \tan\left(R/2\right)}{\mu},
\eea
and $d_J^{q/g, \rm alg}$ represents the algorithm-dependent pieces, 
\bea
d_J^{q,\rm cone} &= \frac{7}{4} + \frac{3}{2}\ln2 - \frac{5\pi^2}{24},
\\
d_J^{q,\rm \text{\kt}} &= \frac{13}{4} - \frac{3\pi^2}{8},
\\
d_J^{g,\rm cone} &=\frac{137}{72} + \frac{11}{6}\ln2 - \frac{5\pi^2}{24} - \frac{T_F n_f}{C_A}\left(\frac{23}{36}+\frac{2}{3}\ln2\right),
\\
d_J^{g,\rm \text{\kt}} & = \frac{67}{18} - \frac{3\pi^2}{8} - \frac{T_Fn_f}{C_A}\frac{23}{18}.
\eea
The unmeasured jet function $J^i(\omega, R, \mu)$ satisfies the RG equation
\bea
\mu\frac{d}{d\mu} J^i(\omega, R, \mu) = \gamma_J^i(\mu) J^i(\omega, R, \mu),
\eea
with the anomalous dimension given as follows:
\bea
\gamma^i_J(\mu) = \Gamma^i_{\rm cusp}(\alpha_s) \ln\frac{\mu^2}{\omega^2 \tan^2(R/2)} + \gamma^i (\alpha_s).
\label{eq:gamma_J}
\eea
Here, $\Gamma^i_{\rm cusp}$ and $\gamma^i$ are the cusp and non-cusp anomalous dimensions, with the perturbative expansions $\Gamma^i_{\rm cusp} = \sum_n \Gamma_{n-1}^i \left(\frac{\alpha_s}{4\pi}\right)^n$ and $\gamma^i = \sum_n \gamma_{n-1}^i \left(\frac{\alpha_s}{4\pi}\right)^n$ where \cite{Becher:2006mr,Becher:2009th,Jain:2011xz,Echevarria:2012pw,Moch:2004pa}
\bea
& \Gamma_0^q = 4 C_F, \qquad  \Gamma_1^q = 4C_F \left[\left(\frac{67}{9}-\frac{\pi^2}{3}C_A - \frac{20}{9}T_F n_f\right)\right], \qquad  \gamma^q_0 = 6C_F,
\\
& \Gamma_{0,1}^g = C_A/C_F \,\Gamma_{0,1}^q, \qquad  \gamma^g_0 = 2\beta_0,
\eea
with $T_F = \frac{1}{2}$, $n_f$ the number of active quark flavors, and
\bea
\beta_0 = \frac{11}{3} C_A - \frac{4}{3} T_F n_f.
\eea
The solution of the RG equation for the unmeasured jet function is
\bea
J^{i}(\omega, R,\mu) = J^i(\omega, R,\mu_J) \exp\left[\int_{\mu_J}^{\mu} \frac{d\mu'}{\mu'} \gamma_J^i(\mu')\right],
\label{eq:J_RGrun}
\eea
where $\mu_J$ is the characteristic scale of $J^{i}(\omega, R,\mu)$,  which eliminates the large logarithms in the fixed-order calculation. From Eqs.~\eqref{eq:Jq} and \eqref{eq:Jg}, the choice of $\mu_J  \sim \omega \tan\left(R/2\right) \equiv p_{TR}$ eliminates the logarithm $L$. We denote this scale as ``$p_{TR}$'' for later convenience. 

\subsection{Fragmenting jet function}
The fragmenting jet functions ${\mathcal G}_i^h(\omega, R, z, \mu)$ \cite{Jain:2011xz,Procura:2011aq,Ritzmann:2014mka} are closely related to the fragmentation functions $D_j^h$ through matching coefficients~${\mathcal J}_{ij}$
\bea
{\mathcal G}_{i}^{h}(\omega, R, z, \mu) = \sum_j \int_z^1 \frac{dx}{x} {\mathcal J}_{ij}\left(\omega, R, x, \mu\right) D_j^h\left(\frac{z}{x}, \mu\right) + {\mathcal O}\left(\frac{\Lambda^2_{\rm QCD}}{\omega^2\tan^2(R/2)}\right),
\label{eq:matching}
\eea
where $D_j^h(z, \mu)$ is the fragmentation function of a parton $j$ fragmenting into a hadron $h$. Eq.~\eqref{eq:matching} is valid for a {\it light} hadron $h$ up to power corrections of order $\Lambda^2_{\rm QCD}/\omega^2\tan^2(R/2)$. Thus, to avoid large non-perturbative power corrections, $R$ should not be too small. On the other hand, for heavy meson fragmenting jet junction $\Lambda_{\rm QCD}$ should be replaced by the heavy quark mass $m_Q$ in the above equation~\cite{Baumgart:2014upa}.

The Wilson coefficients ${\mathcal J}_{ij}$ depend on the jet algorithm. The results for cone jets were given in \cite{Procura:2011aq}, while those for \kt~ jets were first written down in the appendix of \cite{Waalewijn:2012sv}. We provide the detailed derivations of ${\mathcal J}_{ij}$ for \kt~ jets in the appendix, and the results are consistent with \cite{Waalewijn:2012sv}. Here we only list the final results:
\bea
{\mathcal J}_{qq}(\omega, R, z, \mu) &= \delta(1-z) +  \frac{\alpha_s}{\pi} C_F \left[\delta(1-z) \left(L^2 - \frac{\pi^2}{24}\right) + \hat P_{qq}(z) L+ \frac{1-z}{2} + \hat {\mathcal J}_{qq}^{\rm alg}(z) \right],
\label{eq:Jqq}
\\
{\mathcal J}_{qg}(\omega, R, z, \mu) &=  \frac{\alpha_s}{\pi} C_F \left[\frac{z}{2} + \hat P_{gq}(z) L + \hat {\mathcal J}_{qg}^{\rm alg}(z) \right],
\label{eq:Jqg}
\\
{\mathcal J}_{gq}(\omega, R, z, \mu) &= \frac{\alpha_s}{\pi} T_F \left[z (1-z)  + \hat P_{qg}(z) L+\hat {\mathcal J}_{gq}^{\rm alg}(z) \right],
\label{eq:Jgq}
\\
{\mathcal J}_{gg}(\omega, R, z, \mu) &=  \delta(1-z) + \frac{\alpha_s}{\pi} C_A \left[\delta(1-z) \left(L^2 - \frac{\pi^2}{24}\right) + \hat P_{gg}(z) L + \hat {\mathcal J}_{gg}^{\rm alg}(z) \right],
\label{eq:Jgg}
\eea
where the functions $\hat P_{ji}$ have the following expressions~\cite{Jain:2011iu}
\bea
\label{split1}
\hat P_{qq}(z) &= \frac{1+z^2}{(1-z)_+},
\\
\hat P_{gq}(z) &= \frac{1+(1-z)^2}{z},
\\
\hat P_{qg}(z) &= z^2 + (1-z)^2,
\\
\hat P_{gg}(z) &= \frac{2z}{(1-z)_+} + \frac{2(1-z)}{z} + 2z(1-z).
\label{split4}
\eea
$\hat {\mathcal J}_{ij}^{\rm alg}(z)$ represent pieces that depend on the jet algorithm. For cone jets \cite{Procura:2011aq},
\bea
\hat {\mathcal J}_{qq}^{\rm cone} &=
\begin{cases}
  \hat P_{qq}(z) \ln z                                        & z\leq \frac{1}{2} \\
  (1+z^2)\left(\frac{\ln(1-z)}{1-z}\right)_+     & z\geq \frac{1}{2} \\
\end{cases},
\\
\hat {\mathcal J}_{qg}^{\rm cone} &=
\begin{cases}
  \hat P_{gq}(z) \ln z        & z\leq \frac{1}{2}  \\
  \hat P_{gq}(z) \ln(1-z)   & z\geq \frac{1}{2} \\
\end{cases},
\\
\hat {\mathcal J}_{gq}^{\rm cone} &=
\begin{cases}
  \hat P_{qg}(z) \ln z        & z\leq \frac{1}{2}  \\
  \hat P_{qg}(z) \ln(1-z)   & z\geq \frac{1}{2} \\
\end{cases},
\\
\hat {\mathcal J}_{gg}^{\rm cone} &=
\begin{cases}
  \hat P_{gg}(z) \ln z                                                                       & z\leq \frac{1}{2}  \\
  \frac{2(1-z+z^2)^2}{z} \left(\frac{\ln(1-z)}{1-z}\right)_+  & z\geq \frac{1}{2} \\
\end{cases}.
\eea
For \kt~ jets,
\bea
\hat {\mathcal J}_{qq}^{\text{\kt}} &= \hat P_{qq}(z) \ln z+(1+z^2)\left(\frac{\ln(1-z)}{1-z}\right)_+,
\\
\hat {\mathcal J}_{qg}^{\text{\kt}} &= \hat P_{gq}(z) \left(\ln z + \ln(1-z) \right),
\\
\hat {\mathcal J}_{gq}^{\text{\kt}} &= \hat P_{qg}(z) \left(\ln z + \ln(1-z) \right),
\\
\hat {\mathcal J}_{gg}^{\text{\kt}} &= \hat P_{gg}(z) \ln z + \frac{2(1-z+z^2)^2}{z} \left(\frac{\ln(1-z)}{1-z}\right)_+.
\eea
The fragmenting jet function ${\mathcal G}_i^h(\omega, R, z, \mu)$ satisfies the following RG equation
\bea
\mu\frac{d}{d\mu} {\mathcal G}_{i}^{h}(\omega, R, z,\mu) = \gamma_{\mathcal G}^i(\mu) \, {\mathcal G}_{i}^{h}(\omega, R, z,\mu),
\label{eq:gamma_G}
\eea
where the anomalous dimension $\gamma_{\mathcal G}^i(\mu)=\gamma^i_J(\mu)$ is the same as that of the unmeasured jet function $J^i(\omega, R, \mu)$~\cite{Jain:2011xz,Procura:2011aq,Ellis:2010rwa} in Eq.~\eqref{eq:gamma_J}. The solution to the RG equation is
\bea
{\mathcal G}_{i}^{h}(\omega, R, z,\mu) = {\mathcal G}_{i}^{h}(\omega, R, z,\mu_{\mathcal G}) \exp\left[\int_{\mu_{\mathcal G}}^{\mu} \frac{d\mu'}{\mu'} \gamma_{\mathcal G}^i(\mu')\right],
\label{eq:G_RGrun}
\eea
where the scale $\mu_{\mathcal G}$ should be the characteristic scale that eliminates the large logarithms in the fixed-order perturbative calculations. In the large $z$ region, the scale choice $\mu_{\mathcal G} = \omega \tan\left(R/2\right) \left(1-z\right) \equiv p_{TRZ}$ resums \cite{Procura:2011aq} both $\ln R$ and $\ln\left(1-z\right)$. However, for consistency, this would require extracted fragmentation functions $D_{j}^h$ with a built-in resummation of logarithms in $(1-z)$, which is currently not available. It might be instructive to point out that with such a scale, the power corrections in Eq.~\eqref{eq:matching} will be of the order of $\Lambda^2_{\rm QCD}/\left[\omega^2\tan^2(R/2)(1-z)^2\right]$, similar to the usual threshold resummation, see, e.g. Ref.~\cite{Becher:2006mr}. For the numerical calculations presented in the next section, we will choose $\mu_{\mathcal G} = \omega \tan\left(R/2\right)$ to resum $\ln R$ and comment on the effect of $\ln\left(1-z\right)$ resummation.

Let us make a few comments about our resummation formalism. As we have pointed out already at the end of Sec.~\ref{factorization}, since ${\mathcal G}_i^h(\omega, R, z, \mu)$ and $J^i(\omega, R, \mu)$ follow the same RG evolution equations, as given in Eqs.~\eqref{eq:G_RGrun} and \eqref{eq:J_RGrun}, respectively, the ratio $\frac{{\mathcal G}_i^h(\omega, R, z, \mu)}{J^i(\omega, R, \mu)}$ as given in the factorized formalism Eq.~\eqref{eq:master} is thus RG invariant. In other words, this ratio does not depend on the scale $\mu$. Choosing $\mu_{\mathcal G} = \mu_J = \omega \tan(R/2)$, the whole RG exponential forms cancel in the ratio. However, this does not mean that resummation effects disappear in our framework. On the contrary, the resummation effect is shifted entirely into the scale $\mu_{\mathcal G}$-dependence of the standard fragmentation function $D_i^h(z, \mu_{\mathcal G})$ through Eq.~\eqref{eq:matching}. In other words, we are resumming $\ln(R)$ logarithms in this case through the DGLAP evolution equations of the fragmentation functions. This type of resummation was not achieved previously in the fixed NLO calculation of~\cite{Kaufmann:2015hma}. It will be very interesting to explore the exact relation between our work and the previous NLO calculation~\cite{Kaufmann:2015hma}, which we are going to address in a future publication. 

\section{Phenomenology}
\label{sec:phenomenology}
In this section, we present the numerical results of our theoretical formalism and we compare our calculations with the experimental data for both light hadron and heavy meson production at the LHC. We will also explore the theoretical uncertainties of our formalism.

\subsection{Light hadron jet fragmentation function}
We first study the distribution of light hadrons inside jets in proton-proton collisions. Both  ATLAS and  CMS collaborations at the LHC have measured the distribution of light, charged hadrons $h=h^+ + h^-$ inside jets. We perform the numerical calculations using the CT14 NLO parton distribution functions~\cite{Dulat:2015mca} and the DSS07 NLO fragmentation functions~\cite{deFlorian:2007aj,deFlorian:2007hc}. 
We keep the $\Gamma_{0,1}^{i}$ and $\gamma_0^{i}$ terms in the series expansion of the anomalous dimension $\gamma_{J, \mathcal G}^{i}$ with $i=q,g$. Therefore the calculation is at next-to-leading logarithmic  accuracy.

\bef
\includegraphics[width=3.6in]{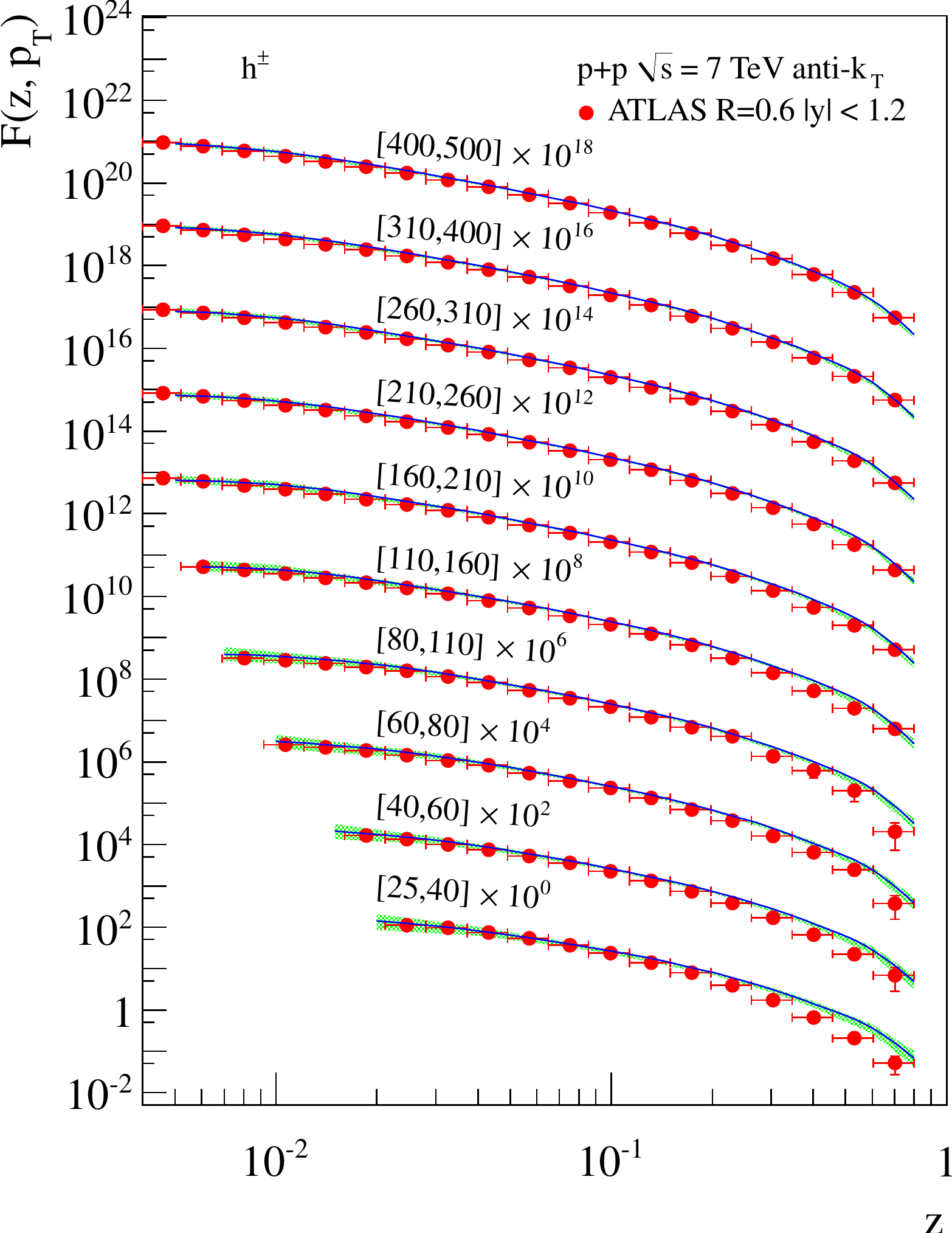}
\caption{Comparison of our theoretical calculations with the ATLAS experimental data~\cite{Aad:2011sc} in proton-proton collisions at $\sqrt{s} = 7$ TeV. Jets are reconstructed using the anti-k$_{\rm T}$~algorithm with $R=0.6$ within the rapidity range $|y|<1.2$. The numbers in the square brackets correspond to different jet transverse momentum bins, e.g. [25, 40] means $25 < p_T < 40$ GeV. The blue solid curves are the ``nominal'' theoretical calculations, where we make the scale choice of $\mu=p_T$, and $\mu_{\mathcal G} = \mu_J = 2p_T\tan\left(R/2\right)\equiv p_{TR}$. The green bands are the estimated uncertainties of our theoretical calculations from scale variations, see the discussion in the text.}
\label{compare-atlas}
\eef

In Fig.~\ref{compare-atlas}, we compare our calculations with the experimental data from ATLAS~\cite{Aad:2011sc} in proton-proton collisions at the center-of-mass (CM) energy of $\sqrt{s} = 7$ TeV. Jets are reconstructed using the anti-k$_{\rm T}$ algorithm with $R=0.6$ within the rapidity range $|y|<1.2$. The transverse momenta $p_T$ of jets are measured across a wide range, from 25~GeV to 500~GeV. The numbers in square brackets correspond to different jet transverse momentum bins, e.g. [25, 40] means $25 < p_T < 40$ GeV. The solid red circles are experimental data, while the solid blue curves are the ``nominal'' theoretical calculations, where we make the scale choices $\mu=p_T$, $\mu_{\mathcal G} =\mu_J = p_{TR}$ defined in the last section. The green bands are the estimated uncertainties of our theoretical calculations from the scale variations for all three scales $\mu,~\mu_{\mathcal G},~\mu_J$ by a factor of 2 around the above central values. See detailed discussions in Sec.~\ref{uncertainty} below. Note that the DSS07 fragmentation function parameterizations for $D_i^h(z, \mu)$ are only valid for $0.05 < z < 1$ and $1<\mu^2< 10^5$ GeV$^2$. Thus, all the calculations outside these regions are based on the extrapolations of the DSS07 parameterizations provided by the distributed package from the authors~\cite{deFlorian:2007aj,deFlorian:2007hc}. As we have expected, the theoretical uncertainties from the scale variations are relatively small, due to the fact that ${\mathcal G}_i^h$ and $J^i$ follow the same RG running as discussed in Sec.~\ref{factorization}. At the same time, as one can see, there is good agreement between our theoretical calculations and the ATLAS data. Our calculations slightly overshoot the experimental data at large $z$ for jets with low $p_T$. Since there are large uncertainties for fragmentation functions in the large $z$ region~\cite{deFlorian:2014xna,Anderle:2015lqa}, jet fragmentation function measurements in proton-proton collisions can help constrain them in this region.

\bef
\includegraphics[width=3.6in]{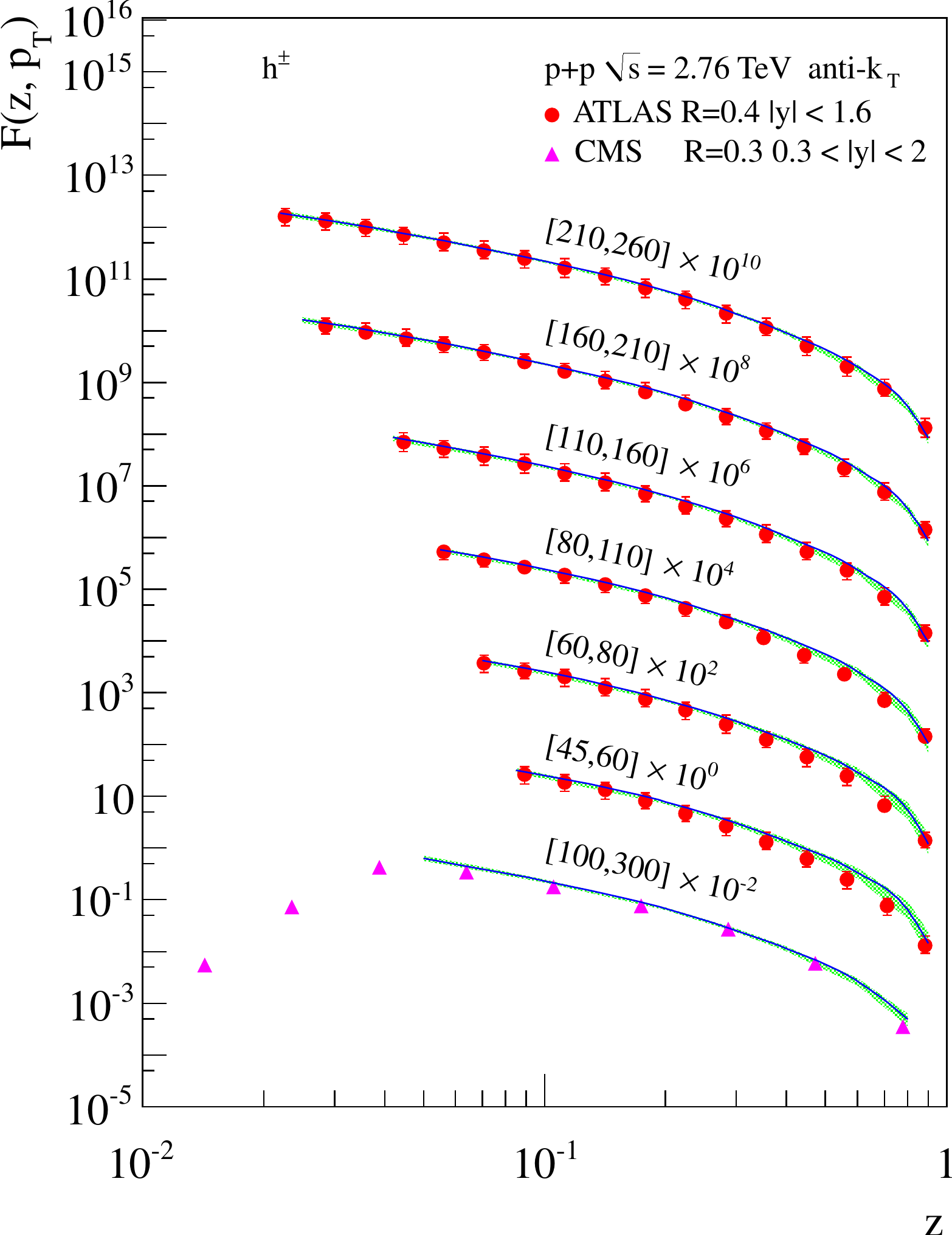}
\caption{Comparison of our theoretical calculations with the LHC data from proton-proton collisions at $\sqrt{s} = 2.76$~TeV. The solid red circles are the ATLAS preliminary data~\cite{ATLAS:2015mla}, while the magenta triangles are the CMS data~\cite{Chatrchyan:2012gw}. The blue solid curves are the ``nominal'' theoretical calculations, with the green bands representing the theoretical uncertainties estimated from scale variations.}
\label{compare-atlas-cms}
\eef

In Fig.~\ref{compare-atlas-cms}, we compare our calculations with the preliminary ATLAS data~\cite{ATLAS:2015mla}, as well as the CMS measurements~\cite{Chatrchyan:2012gw} in proton-proton collisions at the CM energy $\sqrt{s} = 2.76$~TeV. Here, jets are reconstructed using the anti-k$_{\rm T}$ algorithm with $R=0.4$ within the rapidity range $|y|<1.6$ for ATLAS, whereas for CMS $R=0.3$ and $0.3 < |y| < 2$. The solid red circles are the ATLAS data, while the magenta solid triangles are the CMS data. As one can see, our calculations agree with the data rather well. Note that the CMS data has a very different trend for low $z\lesssim 0.05$ compared to the ATLAS data. 
Our theoretical predictions in Figs.~\ref{compare-atlas} and~\ref{compare-atlas-cms} also agree with the results in \cite{Kaufmann:2015hma} that use the full NLO calculation. 

\subsection{Algorithm and radius dependence, and theoretical uncertainty}
\label{uncertainty}
Here, we study the dependence of the jet fragmentation function on the jet algorithm and the jet radius. We will also estimate the theoretical uncertainty by varying the characteristic scales in our formalism.
\bef
\includegraphics[width=3.6in]{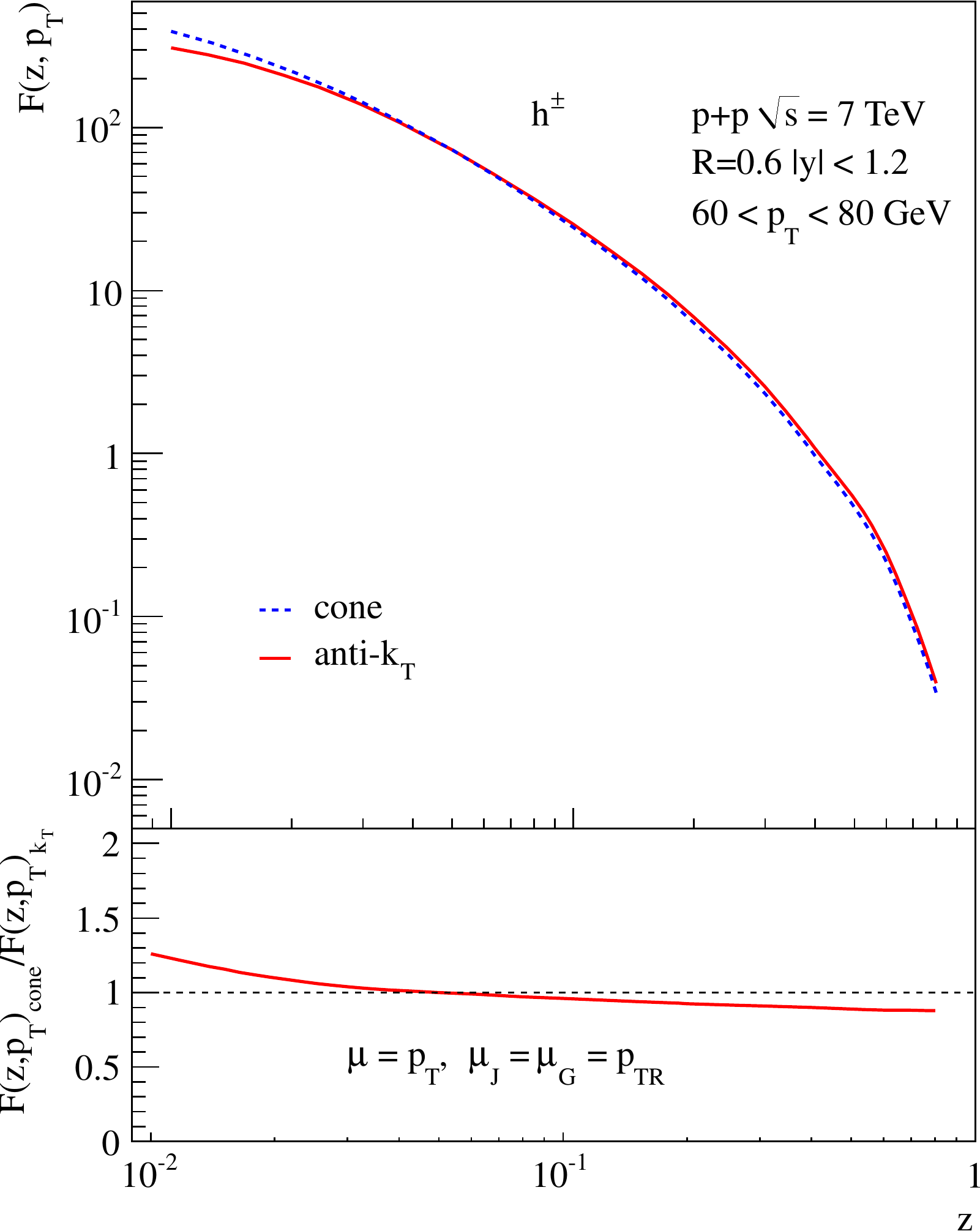}
\caption{Upper panel: Jet fragmentation function $F(z, p_T)$ for light charged hadrons plotted as a function of $z$ for jets with $60<p_T<80$ GeV, $|y| < 1.2$, $R=0.6$) at $\sqrt{s} = 7$ TeV, as an example. We choose the scales $\mu= p_T$ and $\mu_{\mathcal G} = \mu_J = p_{TR}$. The solid red curve is for \kt~ jets, while the dashed blue curve is for cone jets. Lower panel: the ratio of the jet fragmentation functions $F(z, p_T)_{\rm cone}/F(z, p_T)_{\rm k_T}$ for cone and \kt~ jets.}
\label{fig:alg-dependence}
\eef

We will first explore the jet algorithm dependence. In the upper panel of Fig.~\ref{fig:alg-dependence}, we plot the jet fragmentation function $F(z, p_T)$ for light charge hadrons as a function of $z$ inside jets with $60<p_T<80$ GeV, $|y| < 1.2$, $R=0.6$ at $\sqrt{s} = 7$ TeV as an example.
We choose the scales $\mu= p_T$ and $\mu_{\mathcal G} =\mu_J = p_{TR}$. The solid red curve is for \kt~ jets, while the dashed blue curve is for cone jets. As we can see from this plot, $F(z, p_T)$ for cone jets is smaller (larger) than that for \kt~ jets at large (small) $z$. This is a consequence of two combined effects: in the low $z$ region, the FJF ${\mathcal G}_i^h$ for cone jets is larger than that for an \kt~ jet. As $z$ gets closer to 1, the FJF ${\mathcal G}_i^h$ for cone and \kt~ jets approach the same value because there is little radiation left in the jet to distinguish between jet algorithms. Also, the unmeasured jet function $J^i$ for cone jets is larger than that for \kt~ jets. To see the difference more clearly, we plot the ratio $F(z, p_T)_{\rm cone}/F(z, p_T)_{\rm k_T}$ between the jet fragmentation functions for cone and \kt~ jets in the lower panel of Fig.~\ref{fig:alg-dependence}.
\bef
\includegraphics[width=3.6in]{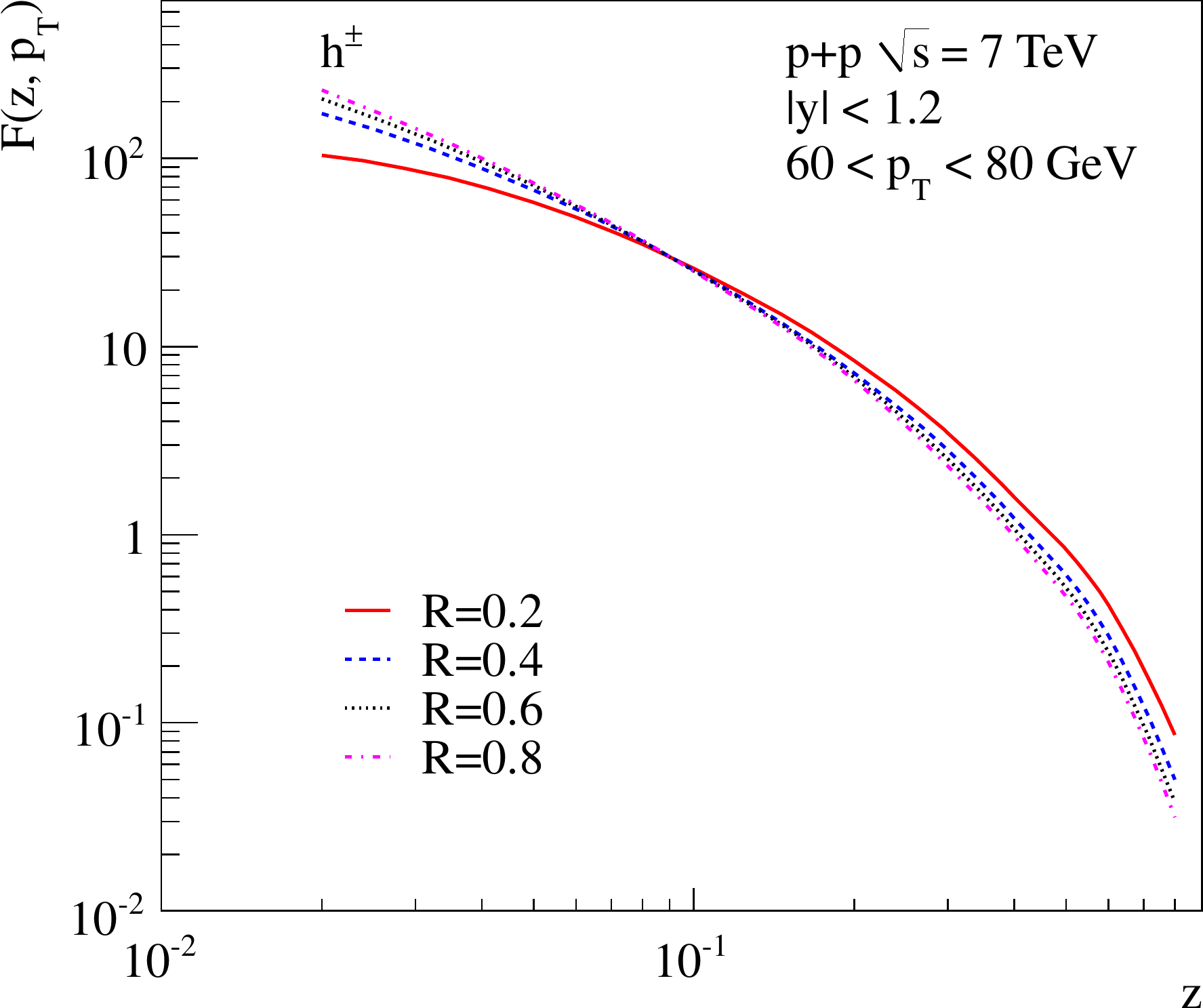}
\caption{Jet fragmentation functions plotted as a function of $z$ for four different jet radii $R=0.2$ (solid red), $R=0.4$ (dashed blue), $R=0.6$ (dotted black), and $R=0.8$ (dash-dotted magenta) for jets with $60<p_T<80$ GeV, $|y| < 1.2$) at $\sqrt{s} = 7$ TeV as an example. We choose the scales $\mu = p_T$ and $\mu_{\mathcal G} =\mu_J = p_{TR}$.} 
\label{fig:R-dependence}
\eef

We now study the jet radius $R$ dependence. We choose the scales $\mu= p_T$ and $\mu_{\mathcal G} =\mu_J = p_{TR}$. In Fig.~\ref{fig:R-dependence}, we plot as an example the jet fragmentation functions $F(z, p_T)$ as a function of $z$ for four different jet radii $R=0.2$ (solid red), $R=0.4$ (dashed blue), $R=0.6$ (dotted black), and $R=0.8$ (dash-dotted magenta) for jets with $60<p_T<80$~GeV, $|y| < 1.2$ at $\sqrt{s} = 7$ TeV. 
We find that in the large $z\gtrsim 0.1$ region $F(z, p_T)$ gets smaller as $R$ increases. In the small $z\lesssim 0.1$ region, $F(z, p_T)$ becomes larger as $R$ increases because of the normalization of $F(z, p_T)$.
This is related to the scale dependence of $D_{i}^h(z, \mu_{\mathcal G})$, which is governed by the DGLAP evolution equations: $D_{i}^h(z, \mu_{\mathcal G})$ increases (decreases) as $\mu_{\mathcal G}$ increases for small (large) $z$~\cite{Field:1989uq}. Since $\mu_{\mathcal G} = p_{TR} = 2p_T \tan(R/2)$, increasing $R$ will increase $\mu_{\mathcal G}$. 
\bef
\includegraphics[width=3.6in]{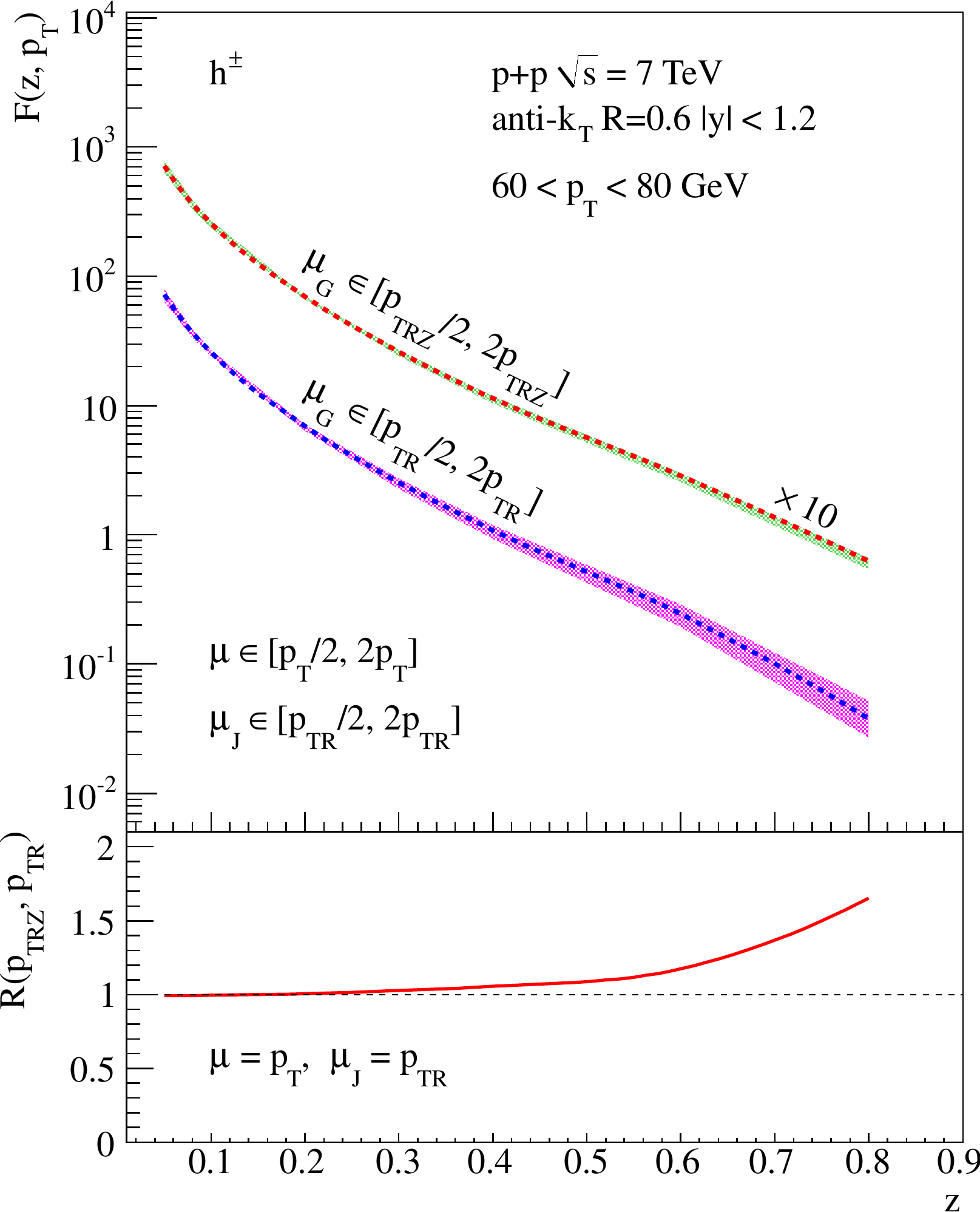}
\caption{Upper panel: Jet fragmentation functions plotted as a function of $z$ for jets with $60<p_T<80$ GeV, $|y| < 1.2$, $R=0.6$ at $\sqrt{s} = 7$ TeV as an example. We vary the scales $\mu\in [p_T/2, 2p_T]$ and $\mu_J\in [p_{TR}/2, 2p_{TR}]$, and for the magenta band $\mu_{\mathcal G}\in [p_{TR}/2, 2p_{TR}]$ while for the green band $\mu_{\mathcal G}\in [p_{TRZ}/2, 2p_{TRZ}]$. Here, $p_{TR} = 2p_T\tan\left(R/2\right)$ and $p_{TRZ} = 2p_T\tan\left(R/2\right) (1-z)$. Lower panel: The ratio $\left.F(z, p_T)\right|_{\mu_{\mathcal G} = p_{_{TRZ}}} /  \left.F(z, p_T)\right|_{\mu_{\mathcal G} = p_{_{TR}}}$ plotted as a function of $z$ with $\mu = p_T$ and $\mu_J = p_{TR}$.}
\label{fig:scale-dependence}
\eef

Finally, we estimate the uncertainty of our theoretical calculations by varying the scales $\mu$, $\mu_J$, and $\mu_{\mathcal G}$ around
\bea
\mu \sim p_T,
\qquad
\mu_J \sim p_{TR} = 2p_T\tan\left(R/2\right),
\qquad
\mu_{\mathcal G} \sim p_{TR} = 2 p_T\tan\left(R/2\right).
\eea
We independently vary the scales by a factor of 2 around their central values, i.e.,
\bea
\mu \in [p_T/2, 2p_T],
\qquad
\mu_J \in [p_{TR}/2, 2 p_{TR}],
\qquad
\mu_{\mathcal G} \in [p_{TR}/2, 2 p_{TR}].
\label{eq:normalvary}
\eea
Thus we have in total 8 different combinations of scales, at which the jet fragmentation function $F(z, p_T)$ is calculated. The uncertainty of $F(z, p_T)$ estimated by the scale variation is then given by the envelope of the results calculated within these regions, i.e. we take the maximum and minimum of the different combinations of the scale variations given in Eq.~\eqref{eq:normalvary} to be the upper and lower boundaries of the uncertainty band. The obtained uncertainty band is shown by the magenta band in the upper panel of Fig.~\ref{fig:scale-dependence}, while the blue dashed curve represents the central value with $\mu=p_T,~\mu_{\mathcal G} = \mu_J = p_{TR}$. The uncertainty of our calculations is generally small for the moderate $z$ region, and it is compatible with the results based on the full NLO calculation in \cite{Kaufmann:2015hma}, where only the variation of the scale $\mu$ is implemented. This gives us confidence that the RG evolutions for both the FJF ${\mathcal G}_i^h$ and the unmeasured jet function $J^i$ indeed improve the convergence of the theoretical calculation.

When $z$ gets closer to 1, one can see that the scale uncertainty band becomes larger. As we have shown in the last section, there is an explicit dependence in the FJF ${\mathcal G}_i^h$ on $\ln(1-z)$. These logarithms become large as $z$ approaches 1, i.e. in the hadronic threshold limit. We may \cite{Procura:2011aq} simultaneously resum logarithms of the jet radius $R$ and $(1-z)$ by choosing the scale $\mu_{\mathcal G} \sim 2p_T \tan\left(R/2\right) \left(1-z\right) \equiv p_{TRZ}$. We plot this by independently varying the scales as follows,
\bea
\mu \in [p_T/2, 2p_T],
\qquad
\mu_J \in [p_{TR}/2, 2 p_{TR}],
\qquad
\mu_{\mathcal G} \in [p_{TRZ}/2, 2 p_{TRZ}].
\label{eq:varylargez}
\eea
Such scale variations correspond to the green band in the upper panel of Fig.~\ref{fig:scale-dependence}, while the red dashed central curve represents the calculation with $\mu=p_T,~\mu_J=p_{TR},~\mu_{\mathcal G} = p_{TRZ}$. As one can clearly see, the uncertainty of the calculation with $\ln(1-z)$ resummation is largely reduced in the large $z$ region.

In order to see the effect of $\ln(1-z)$ resummation more clearly, in the lower panel of Fig.~\ref{fig:scale-dependence}, we plot the ratio $R\left(p_{TRZ}, p_{TR}\right) = \left.F(z, p_T)\right|_{\mu_{\mathcal G} = p_{_{TRZ}}} /  \left.F(z, p_T)\right|_{\mu_{\mathcal G} = p_{_{TR}}}$
as a function of $z$ and we set $\mu=p_T$ and $\mu_J = p_{TR}$. As one can see, resumming $\ln\left(1-z\right)$ leads to an enhancement of the jet fragmentation function $F(z, p_T)$ in the large $z$ region. For $z\gtrsim 0.8$, the enhancement is about a factor of 2. Even though the theoretical uncertainty is reduced with the scale choice $\mu_{\mathcal G} = p_{TRZ}$, we do not use this scale when comparing to data in Figs.~\ref{compare-atlas},~\ref{compare-atlas-cms} above and Fig.~\ref{D-meson} below. This is due to the fact that the fragmentation functions that we use in our numerical studies are extracted using fixed-order calculations~\cite{deFlorian:2007aj,deFlorian:2007hc,deFlorian:2014xna,Anderle:2015lqa}. In order to be consistent, we have to adopt the conventional scale choice $\mu_{\mathcal G} = p_{TR}$. However, we want to make an important point. If one performs a fit for fragmentation functions using the $F(z, p_T)$ data, the extracted functions would differ significantly in the large $z$ region when the more accurate calculation with $\ln\left(1-z\right)$ resummation is used. Our conclusions here are similar to the observations made in~\cite{Anderle:2012rq} in the context of threshold resummation.

\subsection{Heavy meson jet fragmentation function}
Our theoretical result in Eq.~\eqref{eq:master} was derived for light hadron production inside jets. However, it can also be applied to describe  heavy meson production inside jets using the Zero Mass Variable-Flavor Number Scheme (ZMVFNS)~\cite{Collins:1986mp,Stavreva:2009vi}. Such a scheme applies when the perturbative scales $Q$ are much larger than the heavy quark mass $m_Q$: $Q^2\gg m_Q^2$. In this kinematic regime, the heavy quarks are expected to behave like massless partons. One can, thus, treat heavy quarks as the other light partons, and logarithms associated with $m_Q$ are resummed using the DGLAP evolution. Power corrections of ${\mathcal O}(m_Q^2/Q^2)$ are neglected in this formalism. In our case,  the ZMVFNS applies in the kinematic regime where $\mu,~\mu_J,~\mu_{\mathcal G} \gg m_Q$. The ATLAS collaboration has recently measured the distribution of $D^{*\pm}$-mesons in jets with $p_T > 25$ GeV and $R=0.6$~\cite{Aad:2011td}. Given the fact that the charm mass is relatively small $m_c \sim 1.3$ GeV~\cite{Agashe:2014kda}, the jet transverse momentum is large and the radius is moderate, this satisfies the requirement for using the ZMVFNS.

Within the ZMVFNS, the only change in our theoretical formalism is to also include the charm production in Eq.~\eqref{eq:master}: $\sum_{i=q,g,c}$ with $q$ and $c$ representing {\it light} and {\it charm} flavor, respectively. Like in light hadron calculations, we make the scale choices $\mu=p_T$, $\mu_{\mathcal G} =\mu_J = p_{TR}$ for the ``nominal'' calculations. We follow Sec.~\ref{uncertainty} to calculate the theoretical uncertainties from the scale variations. We use the charm-meson fragmentation functions extracted from the inclusive production of a single charm-meson $D$ in $e^+e^-$ collisions: $e^+ e^-\to D\,X$. The parameterizations for $D_{i}^h(z, \mu)$ with $i=q,g,c$ and $h=D$ are available in~\cite{Kneesch:2007ey}, which yield a good description of the inclusive $D$-meson production in proton-proton collisions at the LHC~\cite{Kniehl:2012ti}. Thus, we will use this parametrization in our calculations.

\bef
\includegraphics[width=4.4in]{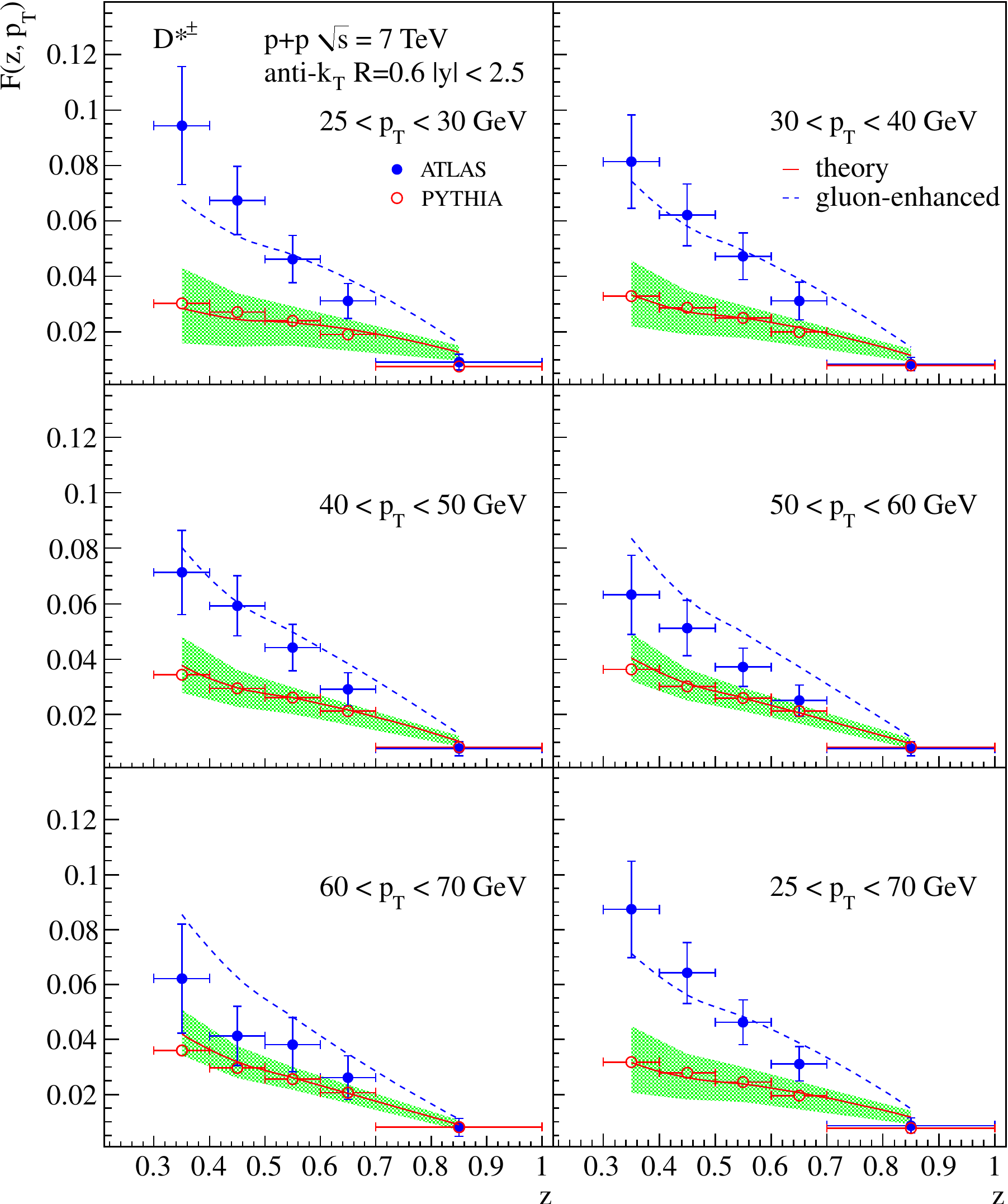}
\caption{The calculation of jet fragmentation functions for $D^{*\pm}$ meson production compared to the experimental data from the ATLAS collaboration at $\sqrt{s}=7$ TeV~\cite{Aad:2011td}. Jets are reconstructed using the anti-k$_{\rm T}$ algorithm with $R=0.6$, and the jet rapidity is within $|y| < 2.5$. We show 6 different panels which correspond to different jet $p_T$ ranges. The solid blue circles are the experimental data measured by ATLAS~\cite{Aad:2011td}, while the empty red circles are the PYTHIA simulations provided in the ATLAS paper~\cite{Aad:2011td}. The solid red curves are our default theoretical calculations using the ZMVFNS. The green bands are the estimated theoretical uncertainties from the scale variations. The dashed blue curves are our calculations using an enhanced gluon-to-$D$ meson fragmentation function: $D_{g}^D(z, \mu) \to 2\,D_{g}^D(z, \mu)$.}
\label{D-meson}
\eef

In Fig.~\ref{D-meson}, we compare our calculations for the $D^{*\pm}$ jet fragmentation function with the ATLAS experimental data at  CM energy $\sqrt{s}=7$ TeV~\cite{Aad:2011td}. Jets are reconstructed using the anti-k$_{\rm T}$ algorithm with $R=0.6$, and the jet rapidity is within $|y| < 2.5$. We show 6 different panels which correspond to different jet $p_T$ ranges covering $25 < p_T < 70$~GeV. The solid blue circles are the experimental data measured by ATLAS~\cite{Aad:2011td} and the empty red circles are the PYTHIA simulations provided in the ATLAS paper~\cite{Aad:2011td}. The solid red curves are our default theoretical calculations, which use the central values of the $D$-meson fragmentation functions $D_{i}^h(z, \mu)$ from~\cite{Kneesch:2007ey}. The green bands are the theoretical uncertainties estimated from the scale variations. As one can clearly see, our theoretical calculations are consistent with the PYTHIA simulations for all different jet $p_T$ bins. However, they are significantly below the experimental measurements from the ATLAS collaboration.

As we have mentioned, the $D$-meson fragmentation functions are extracted in $e^+e^-$ collisions, where the gluon fragmentation function $D_{g}^D(z, \mu)$ does not enter at leading-order in the theoretical formalism. Thus, gluon fragmentation is only indirectly probed through QCD evolution and/or higher-order corrections. This leads to a large uncertainty of the extracted gluon-to-$D$ meson fragmentation function. Note that Ref.~\cite{Kneesch:2007ey} does not provide the uncertainty of the extracted charmed-meson fragmentation functions. However, comparing different extractions from the same sets of $e^+e^-$ data~\cite{Kniehl:2005de,Kniehl:2006mw,Kneesch:2007ey}, we find that the gluon-to-$D$ meson fragmentation function $D_{g}^D(z, \mu)$ can differ by a factor of 3, while quark-to-$D$ meson fragmentation functions $D_{q,c}^D(z, \mu)$ do not vary so dramatically~\cite{Kneesch:2007ey}. Other than that, the various extractions~\cite{Kniehl:2005de,Kniehl:2006mw,Kneesch:2007ey} differ only by the initial scales for the QCD evolution or by the treatment of the heavy quark mass. This provides a strong hint that the current extraction of the gluon-to-$D$ meson fragmentation function could have a very large uncertainty.

To explore the uncertainty of the gluon-to-$D$ meson fragmentation function, we re-perform our calculations of the jet fragmentation functions for $D^{*\pm}$ meson with the gluon-to-$D$ meson fragmentation function enhanced by a factor of 2, i.e. $D_{g}^D(z, \mu) \to 2\,D_{g}^D(z, \mu)$. These calculations are shown by the dashed blue curves in Fig.~\ref{D-meson}. They lead to much better agreement with the ATLAS data. We have also tried enhancing other quark-to-$D$ meson fragmentation functions $D_{q,c}^D(z, \mu)$ by a similar factor, but none of them could lead to such an efficient enhancement in the jet fragmentation function. We conclude that jet fragmentation functions of heavy mesons in proton-proton collisions have great potential to constrain the gluon-to-heavy meson fragmentation functions. 

\section{Summary}
\label{sec:summary}
In this paper, we studied  jet fragmentation functions for light hadrons and heavy mesons inside reconstructed jets. We wrote down a factorized expression in SCET for the jet fragmentation function in proton-proton collisions. We found that, up to power corrections, the jet fragmentation function can be expressed as the ratio of the fragmenting jet function and the unmeasured jet function. These two functions satisfy the same renormalization group equation, and the fragmenting jet function can be further expressed as a convolution between the fragmentation functions and the matching coefficients. Using SCET, we were able to simultaneously resum large logarithms of the jet radius $R$ and $(1-z)$, which has a significant impact on the phenomenology considered in this work. We used the theoretical formalism to describe the jet fragmentation functions for light hadron and heavy meson production measured at the LHC. We found that our calculations agree very well with the experimental data for light hadron production. We explored the jet algorithm and the $R$ dependence of the jet fragmentation functions, and we estimated the theoretical uncertainty by scale variation. For heavy meson production inside jets, although our calculations are consistent with PYTHIA simulations, they fail to describe the corresponding LHC data. We found that enhancing the gluon-to-heavy meson fragmentation function leads to much better agreement with the experimental data. We emphasize that the jet fragmentation function for heavy meson production in proton-proton collisions is very sensitive to the gluon-to-heavy meson fragmentation function. In the future, we plan to extend our calculations to describe jet fragmentation functions in heavy ion collisions in order to understand nuclear modifications of hadron production inside jets.

\acknowledgments
We would like to thank Emanuele Mereghetti for his collaboration at the early stage of the project, and for many useful discussions and comments throughout the project. We also thank Andrew Hornig, Daekyoung Kang, Christopher Lee, Adam Leibovich, Thomas Mehen, Werner Vogelsang, and Wouter Waalewijn for helpful discussions and comments. This work is supported by the U.S. Department of Energy under Contract No.~DE-AC52-06NA25396, 
the Office of Science Early Career Award program,  and in part by the LDRD program at Los Alamos National Laboratory.

\appendix
\section{Matching calculations}
\label{sec:appendix}
Here, we derive the matching coefficients ${\mathcal J}_{ij}$ in the fragmenting jet function ${\mathcal G}_i^h(\omega, R, z, \mu) $ with the fragmentation function $D_i^h(z, \mu)$ for \kt~ jets. These results were first written down in the appendix of \cite{Waalewijn:2012sv}. Here we provide the detailed derivations of ${\mathcal J}_{ij}$ for \kt~ jets, and the results are consistent with \cite{Waalewijn:2012sv}. We start by specifying the phase space constraint from the jet algorithm, which was nicely discussed in~\cite{Ellis:2010rwa}. Consider a parton splitting process, $i(\ell)\to j(q)+k(\ell - q)$, where an incoming parton $i$ with momentum $\ell$ splits into a parton $j$ with momentum $q$ and a parton $k$ with momentum $\ell-q$. The four-vector $\ell^\mu$ can be decomposed in light-cone coordinates as $\ell^\mu=(\ell^+, \ell^- = \omega, 0_\perp)$ where $\ell^{\pm} = \ell^0\mp \ell^z$. The constraints for cone and \kt~ algorithms with radius $R$ are given by
\bea
\text{cone:~~} & \Theta_{\rm cone} =  \theta\left(\tan^2\frac{R}{2} - \frac{q^+}{q^-}\right)  \theta\left(\tan^2\frac{R}{2} - \frac{\ell^+ - q^+}{\omega - q^-}\right),
\\
\text{\kt:~~} & \Theta_{\text{\kt}} =   \theta\left(\tan^2\frac{R}{2} - \frac{q^+\omega^2}{q^-\left(\omega - \ell^-\right)^2}\right).
\eea

For fragmenting jet functions, the above constraints lead to constraints on the jet invariant mass $m_J^2 = \omega \ell^+$~\cite{Procura:2011aq}, which are derived and listed as follows:
\bea
\text{cone:~~} & \delta_{\rm cone} = \theta\left({\rm min}\left(\frac{z}{1-z},~\frac{1-z}{z}\right) \omega^2\tan^2\frac{R}{2} - m_J^2 \right) \theta(m_J^2),
\label{eq:cone}
\\
\text{\kt:~~} &\delta_{\text{\kt}} =  \theta\left( z (1-z) \omega^2\tan^2\frac{R}{2} - m_J^2 \right) \theta(m_J^2),
\label{eq:kt}
\eea
with $z=q^-/\omega$. The FJF ${\mathcal G}_i^h(\omega, R, z, \mu)$ is therefore related to the fragmenting jet functions ${\mathcal G}_{i}^{h}(m_J^2, z, \mu)$~\cite{Jain:2011xz} with the extra measurement of jet mass through
\bea
{\mathcal G}_i^h(\omega, R, z, \mu) = \int dm_J^2\, {\mathcal G}_{i}^{h}(m_J^2, z, \mu)\, \delta_{\rm alg},
\label{eq:mJ}
\eea
where $\delta_{\rm alg} = \delta_{\rm cone}$ or $\delta_{\text{\kt}}$ are the constraints given in Eqs.~\eqref{eq:cone} and \eqref{eq:kt}. The FJF ${\mathcal G}_i^h(\omega, R, z, \mu)$ can be matched onto the fragmentation function $D_i^h(z, \mu)$:
\bea
{\mathcal G}_{i}^{h}(\omega, R, z, \mu) = \sum_j \int_z^1 \frac{dx}{x} {\mathcal J}_{ij}\left(\omega, R, x, \mu\right) D_j^h(\frac{z}{x}, \mu) + {\mathcal O}\left(\frac{\Lambda^2_{\rm QCD}}{\omega^2\tan^2(R/2)}\right),
\label{eq:relation}
\eea
and ${\mathcal J}_{ij}$ are the matching coefficients. 

The FJF ${\mathcal G}_{i}^{j}(m_J^2, z, \mu)$ with $i,j=q,g$ has been extensively studied in \cite{Jain:2011xz,Ritzmann:2014mka}. Using pure dimensional regularization with $4-2\epsilon$ dimensions in the $\overline{\rm MS}$ scheme, the bare results at ${\cal O}(\alpha_s)$ can be written in the following compact form~\cite{Giele:1991vf,Ritzmann:2014mka}:
\bea
{\mathcal G}_{i, {\rm bare}}^j(m_J^2, z) = \frac{\alpha_s}{2\pi} \frac{\left(e^{\gamma_E}\mu^2\right)^\epsilon}{\Gamma(1-\epsilon)} P_{ji}(z, \epsilon) z^{-\epsilon} (1-z)^{-\epsilon} \left(m_J^2\right)^{-1-\epsilon},
\label{eq:original}
\eea
where the functions $P_{ji}(z, \epsilon)$ are 
\bea
P_{qq}(z, \epsilon) &= C_F\left[\frac{1+z^2}{1-z} - \epsilon\, (1-z) \right],
\\
P_{gq}(z, \epsilon) &= C_F\left[\frac{1+(1-z)^2}{z} - \epsilon\, z \right],
\\
P_{qg}(z, \epsilon) &= T_F\left[1-\frac{2z(1-z)}{1-\epsilon}\right],
\\
P_{gg}(z, \epsilon) &= C_A\left[\frac{2z}{1-z} + \frac{2(1-z)}{z} + 2z(1-z) \right].
\eea
Substituting  Eq.~\eqref{eq:original} into Eq.~\eqref{eq:mJ} and performing the integration over $m_J^2$ with the constraints imposed by the jet algorithm $\delta_{\rm alg}$, one obtains the bare FJF ${\mathcal G}_{i,\rm bare}^j(\omega, R, z)$. We present the results for \kt~ jets here, as the explicit expressions are not available in the literature:
\bea
{\mathcal G}_{q, {\rm bare}}^{q}(\omega, R, z) =& \frac{\alpha_s}{2\pi} C_F\left[\frac{1}{\epsilon^2} +\frac{3}{2\epsilon}-\frac{2}{\epsilon} L \right] \delta(1-z)
\nnu
& + \frac{\alpha_s}{2\pi} C_F \left(-\frac{1}{\epsilon}\right) \left[\hat P_{qq}(z) + \frac{3}{2}\delta(1-z) \right]
\nnu
& + \frac{\alpha_s}{\pi} C_F \Bigg[\delta(1-z) \left(L^2 - \frac{\pi^2}{24}\right) + \hat P_{qq}(z) \left(L+\ln z\right)
\nnu
&+ \frac{1-z}{2} + (1+z^2)\left(\frac{\ln(1-z)}{1-z}\right)_+\Bigg],
\label{eq:qq}
\\
{\mathcal G}_{q, {\rm bare}}^{g}(\omega, R, z) =&  \frac{\alpha_s}{2\pi} C_F \left(-\frac{1}{\epsilon}\right) \hat P_{gq}(z) +  \frac{\alpha_s}{\pi} C_F \left[\frac{z}{2} + \hat P_{gq}(z) \left(L+\ln z + \ln(1-z) \right) \right],
\label{eq:gq}
\\
{\mathcal G}_{g, {\rm bare}}^{q}(\omega, R, z) =&  \frac{\alpha_s}{2\pi} T_F \left(-\frac{1}{\epsilon}\right) \hat P_{qg}(z) + \frac{\alpha_s}{\pi} T_F \left[z (1-z)  + \hat P_{qg}(z) \left(L+\ln z + \ln(1-z) \right) \right],
\label{eq:qg}
\\
{\mathcal G}_{g, {\rm bare}}^{g}(\omega, R, z) =& \frac{\alpha_s}{2\pi} C_A\left[\frac{1}{\epsilon^2} +\frac{1}{\epsilon} \frac{\beta_0}{2C_A} - \frac{2}{\epsilon} L \right] \delta(1-z)
\nnu
& + \frac{\alpha_s}{2\pi} C_A \left(-\frac{1}{\epsilon}\right) \left[\hat P_{gg}(z) + \frac{\beta_0}{2C_A}\delta(1-z) \right]
\nnu
& + \frac{\alpha_s}{\pi} C_A \Bigg[\delta(1-z) \left(L^2 - \frac{\pi^2}{24}\right) + \hat P_{gg}(z) \left(L+\ln z\right)
\nnu
&+ \frac{2(1-z+z^2)^2}{z} \left(\frac{\ln(1-z)}{1-z}\right)_+ \Bigg],
\label{eq:gg}
\eea
where, as a reminder,  $\beta_0$ and $L$ are given by
\bea
\beta_0 = \frac{11}{3} C_A - \frac{4}{3} T_F n_f,
\qquad
L = \ln\frac{\omega \tan\left(R/2\right)}{\mu},
\eea
and $\hat P_{ji}$ have the  expressions~\cite{Jain:2011iu} given in Eq.~(\ref{split1}) - (\ref{split4}).
It is instructive to point out that the $\epsilon$ poles in the first line of Eqs.~\eqref{eq:qq} and \eqref{eq:gg} correspond to ultraviolet (UV) divergences, and they are related to the renormalization of the FJF ${\mathcal G}_{i, \rm bare}^{j}$. All the remaining $\epsilon$ poles in Eqs.~(\ref{eq:qq} - \ref{eq:gg}) are infrared (IR), and they match exactly those in the fragmentation functions $D_{i}^j$ as we will show below.

The renormalization of ${\mathcal G}_{i, {\rm bare}}^{h}(\omega, R, z)$ is given by
\bea
{\mathcal G}_{i, {\rm bare}}^{h}(\omega, R, z) = Z_{\mathcal G}^i(\mu)\, {\mathcal G}_{i}^{h}(\omega, R, z,\mu),
\eea
where $i$ is not summed over on the right hand side. The corresponding renormalization group (RG) equation is given by
\bea
\mu\frac{d}{d\mu} {\mathcal G}_{i}^{h}(\omega, R, z,\mu) = \gamma_{\mathcal G}^i(\mu) \, {\mathcal G}_{i}^{h}(\omega, R, z,\mu),
\label{eq:RG}
\eea
where the anomalous dimension $\gamma_{\mathcal G}^i(\mu)$ is
\bea
 \gamma_{\mathcal G}^i(\mu) = - \left(Z_{\mathcal G}^i(\mu)\right)^{-1} \mu \frac{d}{d\mu} Z_{\mathcal G}^i(\mu).
\eea
The solution to Eq.~\eqref{eq:RG} is then
\bea
{\mathcal G}_{i}^{h}(\omega, R, z,\mu) = {\mathcal G}_{i}^{h}(\omega, R, z,\mu_{\mathcal G}) \exp\left[\int_{\mu_{\mathcal G}}^{\mu} \frac{d\mu'}{\mu'} \gamma_{\mathcal G}^i(\mu')\right],
\label{eq:Grun}
\eea
where the scale $\mu_{\mathcal G}$ should be the characteristic scale chosen such that large logarithms in the fixed-order calculation are eliminated. The counter terms $Z_{\mathcal G}^i(\mu)$ are given by
\bea
Z_{\mathcal G}^q (\mu) &= 1 + \frac{\alpha_s}{2\pi} C_F\left[\frac{1}{\epsilon^2} +\frac{3}{2\epsilon}-\frac{2}{\epsilon} L \right],
\label{eq:uvq}
\\
Z_{\mathcal G}^g (\mu) &= 1 +  \frac{\alpha_s}{2\pi} C_A\left[\frac{1}{\epsilon^2} +\frac{1}{\epsilon} \frac{\beta_0}{2C_A} - \frac{2}{\epsilon} L \right].
\label{eq:uvg}
\eea
From these results we obtain the anomalous dimension $\gamma_{\mathcal G}^i(\mu)$ with the following form,
\bea
\gamma_{\mathcal G}^i (\mu) = \Gamma^i_{\rm cusp}(\alpha_s) \ln\frac{\mu^2}{\omega^2 \tan^2(R/2)} + \gamma^i (\alpha_s),
\eea
where $\Gamma^i_{\rm cusp} = \sum_n \Gamma_{n-1}^i \left(\frac{\alpha_s}{4\pi}\right)^n$ and $\gamma^i = \sum_n \gamma_{n-1}^i \left(\frac{\alpha_s}{4\pi}\right)^n$. The lowest-order coefficients can be extracted from the above calculations:
\bea
 \Gamma_0^q = 4 C_F, \qquad & \gamma^q_0 = 6C_F,
 \\
 \Gamma_0^g = 4 C_A, \qquad & \gamma^g_0 = 2\beta_0,
\eea
and higher-order results can be found in~\cite{Becher:2006mr,Becher:2009th,Jain:2011xz,Echevarria:2012pw,Moch:2004pa}.

After the subtraction of the UV counter terms specified in Eqs.~\eqref{eq:uvq} and \eqref{eq:uvg}, the renormalized FJF ${\mathcal G}_{i}^{j}(\omega, R, z,\mu)$ are given by
\bea
{\mathcal G}_{q}^{q}(\omega, R, z, \mu) =& \frac{\alpha_s}{2\pi} C_F \left(-\frac{1}{\epsilon}\right) \left[\hat P_{qq}(z) + \frac{3}{2}\delta(1-z) \right]
\nnu
& + \frac{\alpha_s}{\pi} C_F \Bigg[\delta(1-z) \left(L^2 - \frac{\pi^2}{24}\right) + \hat P_{qq}(z) \left(L+\ln z\right)
\\
&+ \frac{1-z}{2}+(1+z^2)\left(\frac{\ln(1-z)}{1-z}\right)_+\Bigg],
\label{eq:qqR}
\\
{\mathcal G}_{q}^{g}(\omega, R, z, \mu) =&  \frac{\alpha_s}{2\pi} C_F \left(-\frac{1}{\epsilon}\right) \hat P_{gq}(z) +  \frac{\alpha_s}{\pi} C_F \left[\frac{z}{2} + \hat P_{gq}(z) \left(L+\ln z + \ln(1-z) \right) \right],
\label{eq:gqR}
\\
{\mathcal G}_{g}^{q}(\omega, R, z, \mu) =&  \frac{\alpha_s}{2\pi} T_F \left(-\frac{1}{\epsilon}\right) \hat P_{qg}(z) + \frac{\alpha_s}{\pi} T_F \left[z (1-z)  + \hat P_{qg}(z) \left(L+\ln z + \ln(1-z) \right) \right],
\label{eq:qgR}
\\
{\mathcal G}_{g}^{g}(\omega, R, z, \mu) =& \frac{\alpha_s}{2\pi} C_A \left(-\frac{1}{\epsilon}\right) \left[\hat P_{gg}(z) + \frac{\beta_0}{2C_A}\delta(1-z) \right]
\nnu
& + \frac{\alpha_s}{\pi} C_A \Bigg[\delta(1-z) \left(L^2 - \frac{\pi^2}{24}\right) + \hat P_{gg}(z) \left(L+\ln z\right)
\nnu
&+ \frac{2(1-z+z^2)^2}{z} \left(\frac{\ln(1-z)}{1-z}\right)_+ \Bigg],
\label{eq:ggR}
\eea
where we can eliminate all large logarithms $L$ by choosing $\mu = \omega \tan(R/2)$. 

At the intermediate scale $\mu_{\mathcal G}\gg \Lambda_{\rm QCD}$, one can match the FJF ${\mathcal G}_{i}^{h}(\omega, R, z, \mu)$ onto the fragmentation function $D_j^h(z, \mu)$ as in Eq.~\eqref{eq:relation}. In order to perform the matching calculation and determine the coefficients ${\mathcal J}_{ij}$, we simply need the perturbative results for the fragmentation function for a parton $i$ fragmenting into a parton $j$, $D_i^j(x, \mu)$.
The renormalized $D_i^j(x, \mu)$ at ${\cal O}(\alpha_s)$ using pure dimensional regularization are given by
\bea
D_{q}^q(x, \mu) &= \delta(1-x) + \frac{\alpha_s}{2\pi} C_F \left(-\frac{1}{\epsilon}\right) \left[\hat P_{qq}(x) + \frac{3}{2}\delta(1-x) \right],
\\
D_{q}^g(x, \mu) &= \frac{\alpha_s}{2\pi} C_F \left(-\frac{1}{\epsilon}\right) \hat P_{gq}(x),
\\
D_{g}^q(x, \mu) &= \frac{\alpha_s}{2\pi} T_F \left(-\frac{1}{\epsilon}\right) \hat P_{qg}(x),
\\
D_{g}^g(x, \mu) &= \delta(1-x) + \frac{\alpha_s}{2\pi} C_A \left(-\frac{1}{\epsilon}\right) \left[\hat P_{gg}(x) + \frac{\beta_0}{2C_A}\delta(1-x) \right].
\eea
Using the results for both ${\mathcal G}_{i}^{j}(\omega, R, z, \mu)$ and $D_i^j(x, \mu)$, we obtain the following matching coefficients:
\bea
{\mathcal J}_{qq}(\omega, R, z, \mu) &= \delta(1-z) +  \frac{\alpha_s}{\pi} C_F \left[\delta(1-z) \left(L^2 - \frac{\pi^2}{24}\right) + \hat P_{qq}(z) L+ \frac{1-z}{2} + \hat {\mathcal J}_{qq}^{\rm alg}(z) \right],
\\
{\mathcal J}_{qg}(\omega, R, z, \mu) &=  \frac{\alpha_s}{\pi} C_F \left[\frac{z}{2} + \hat P_{gq}(z) L + \hat {\mathcal J}_{qg}^{\rm alg}(z) \right],
\\
{\mathcal J}_{gq}(\omega, R, z, \mu) &= \frac{\alpha_s}{\pi} T_F \left[z (1-z)  + \hat P_{qg}(z) L+\hat {\mathcal J}_{gq}^{\rm alg}(z) \right],
\\
{\mathcal J}_{gg}(\omega, R, z, \mu) &=  \delta(1-z) + \frac{\alpha_s}{\pi} C_A \left[\delta(1-z) \left(L^2 - \frac{\pi^2}{24}\right) + \hat P_{gg}(z) L + \hat {\mathcal J}_{gg}^{\rm alg}(z) \right],
\eea
where $\hat {\mathcal J}_{ij}^{\rm alg}(z)$ are jet-algorithm dependent. For \kt~jets,
\bea
\hat {\mathcal J}_{qq}^{\text{\kt}} &= \hat P_{qq}(z) \ln z+(1+z^2)\left(\frac{\ln(1-z)}{1-z}\right)_+,
\\
\hat {\mathcal J}_{qg}^{\text{\kt}} &= \hat P_{gq}(z) \left(\ln z + \ln(1-z) \right),
\\
\hat {\mathcal J}_{gq}^{\text{\kt}} &= \hat P_{qg}(z) \left(\ln z + \ln(1-z) \right),
\\
\hat {\mathcal J}_{gg}^{\text{\kt}} &= \hat P_{gg}(z) \ln z + \frac{2(1-z+z^2)^2}{z} \left(\frac{\ln(1-z)}{1-z}\right)_+,
\eea
which are consistent with those from \cite{Waalewijn:2012sv}, whereas the results for cone jets are available in~\cite{Procura:2011aq}.

Substituting the matching coefficients ${\mathcal J}_{ij}$ into Eq.~\eqref{eq:relation}, and writing out explicitly the plus functions, one obtains
\bea
{\mathcal G}_q^h(\omega, R, z, \mu) &=\left\{1+\frac{\alpha_s}{\pi} C_F  \left[ \ln^2\left(\frac{\omega \tan(R/2) (1-z)}{\mu}\right) - \frac{\pi^2}{24}\right] \right\} D_q^h(z, \mu) + \cdots\, ,
\\
{\mathcal G}_g^h(\omega, R, z, \mu) &= \left\{1+\frac{\alpha_s}{\pi} C_A  \left[ \ln^2\left(\frac{\omega \tan(R/2) (1-z)}{\mu}\right) - \frac{\pi^2}{24}\right] \right\} D_g^h(z, \mu) + \cdots\, ,
\eea
where the ellipses represent terms which are regular as $z\to 1$. In the large $z\to 1$ region, there are additional logarithms $\sim \ln(1-z)$. One may choose the scale $\mu = \omega \tan(R/2) (1-z)$ and simultaneously resum both logarithms of $R$ and $(1-z)$ \cite{Procura:2011aq}. The numerical results of this scale choice compared to those by choosing $\mu = \omega \tan(R/2)$ are discussed in section~\ref{sec:phenomenology}.

\providecommand{\href}[2]{#2}\begingroup\raggedright\endgroup

\end{document}